# A negative dielectric constant in nano-particle materials under an electric field at very low frequencies

C. W. Chu[1,2,3], F. Chen[1], J. Shulman[1], S. Tsui[1], Y. Y. Xue[1], W. Wen[4] and P. Sheng[4]

[1]Department of Physics and Texas Center for Superconductivity, University of Houston, 202 Houston Science Center, Houston, Texas 77204-5002
[2]Hong Kong University of Science and Technology, Clear Water Bay, Kowloon, Hong Kong
[3]Lawrence Berkeley National Laboratory, 1 Cyclotron Road, Berkeley, California 94720
[4]Department of Physics and Institute of Nano Science and Technology, Hong Kong University of Science and Technology, Clear Water Bay, Kowloon, Hong Kong

## ABSTRACT

The significance of a negative dielectric constant has long been recognized. We report here the observation of a field-induced large negative dielectric constant of aggregates of oxide nano-particles at frequencies below ~ 1 Hz at room temperature. The accompanying induced charge detected opposes the electric field applied in the field-induced negative dielectric constant state. A possible collective effect in the nano-particle aggregates is proposed to account for the observations. Materials with a negative dielectric constant are expected to provide an attraction between similar charges and unusual scattering to electromagnetic waves with possible profound implications for high temperature superconductivity and communications.

## 1. A NEGATIVE STATIC DIELECTRIC CONSTANT

According to the BCS theory, the superconducting transition temperature $T_c$ is equal to $1.14\theta_D \exp[-1/N(0)V]$, where $\theta_D$ is the Debye temperature (or the characteristic temperature of other quasi-particles), $N(0)$ is the electron density of states at the Fermi energy, and $V$ is the attractive interaction between the electrons (or other quasi-particles). It is evident that the greater the $V$ is, the higher the $T_c$ is. Various propositions have been advanced over the years prior to the discovery of cuprate high temperature superconductivity to enhance $V$ and thus $T_c$.[1] The main interaction between two electric charges, $q_1$ and $q_2$, separated by a distance $r$, is known to be $V(r) = q_1 q_2 / \varepsilon' r$, where $\varepsilon'$ is the real part of the dielectric constant $\varepsilon = \varepsilon' - i\varepsilon''$. A negative static dielectric constant, $\varepsilon'(k,\omega)$ at a wave-vector $k \neq 0$ and frequency $\omega = 0$, has attracted great interest because of its implications for wave propagation, electrolyte behavior, colloidal properties, and bio-membrane functions in addition to possible high temperature superconductivity. For a negative $\varepsilon'$, the force between two electrons will be attractive, leading to electron-pairing if other electron instabilities can be arrested.[2] Electron-pairing has been demonstrated to be responsible for superconductivity in the conventional low temperature superconductors as well as in the newly discovered high temperature superconductors. A route via a negative $\varepsilon'$ to higher $T_c$ may be worth exploring.

The analogy between magnetism and electricity has long been established by Maxwell in the 19th century, in spite of their subtle difference. While magnetic materials display paramagnetism, ferromagnetism, antiferromagnetism, and diamagnetism, only paraelectricity, ferroelectricity, and antiferrolelectricity have been found in dielectric materials. The missing "diaelectricity" may therefore be found in a dielectric material that is characterized by a negative dielectric susceptibility $\varepsilon'-1$ and a polarization opposing the electric field.

For the above reasons, we have decided to examine materials that may display a negative $\varepsilon'$.

## 2. THE EXISTENCE OF A NEGATIVE STATIC DIELECTRIC CONSTANT

Extensive debates concerning the existence of a negative dielectric constant have been carried out for decades based on the consideration of causality and instabilities.[2] The consequence (C) of an electric system subjected to a controllable external cause (A) is related to the response function (R) as C ≡ R x A. The Kramers-Kronig relations state that the Fourier components of the real and imaginary parts of R(k, ω) = R'(k, ω) + $\iota$R"(k, ω) are related through a Hilbert integral transformation, *e.g.*

$$R'(k, \omega) = 1 + \frac{2}{\pi}\int_0^\infty \frac{\omega' d\omega'}{\omega'^2 - \omega^2} R''(k, \omega'), \qquad (1)$$

and in the static limit,

$$R'(k, 0) = 1 + \frac{2}{\pi}\int_0^\infty \frac{\omega' d\omega'}{\omega'^2} R''(k, \omega'). \qquad (2)$$

Discussions based on the Kramers-Kronig relations for R invariably lead to a conclusion that a negative ε' is not admissible when ε' is taken as R'. However, The Maxwell equations relating the displacement D and the electric field E with the external charge density $\rho_{ext}$ and the total charge density $\rho_{tot} = \rho_{ext}$ + the induced charge density are

$$\nabla \cdot D(k, \omega) = 4\pi\rho_{ext}(k, \omega), \text{ and} \qquad (3)$$

$$\nabla \cdot E(k, \omega) = 4\pi\rho_{tot}(k, \omega). \qquad (4)$$

Since D(k, ω) ≡ ε'(k, ω)E(k, ω), $\rho_{ext}$(k, ω) = ε'(k, ω)$\rho_{tot}$(k, ω). Because $\rho_{ext}$(k, ω) is the controllable external cause, it is only logical to identify $\rho_{ext}$(k, ω) as the external cause A and $\rho_{tot}$(k, ω) the consequence C with 1/ε'(k, 0) as the response function R.[3] According to C ≡ R x A, the proper causality relation is then

$$\rho_{tot}(k, \omega) = [1/\varepsilon(k, \omega)]\rho_{ext}(k, \omega). \qquad (5)$$

The Kramers-Kronig relation now becomes

$$1/\varepsilon'(k, 0) = 1 + \frac{2}{\pi}\int \frac{\omega' d\omega'}{\omega'^2} [1/\varepsilon''(k, \omega')]. \qquad (6)$$

It follows[2] that 1/ε'(k, 0) ≤ 1 or equivalently ε'(k, 0) ≥ 1 and ε'(k, 0) ≤ 0. One therefore concludes that ε'(k, 0) can be negative or greater than or equal to +1, assuming that instabilities leading to a possible phase separation can be avoided. A state of negative compressibility in 2D carrier gasses has recently been shown to survive such a phase separation under rather broad conditions.[4]

## 3. THE ANOMALOUS DIELECTRIC BEHAVIOR OF SAMPLES OF NANO-PARTICLES OR WITH NANO-STRUCTURES

In spite of the conclusion given above that a static negative ε'(k,0) is possible, none has been reported. However, artificial nano-structures have been proposed to modify the plasma parameters and to achieve a negative ε'(k,ω),[5,6] albeit in the microwave region. Recent experiments[7] did confirm the prediction. An unusual ε'-dispersion has been detected in samples of nano-particles or with nano-defects at low frequencies, resulting in a very large ε'.[8,9] A giant electric-field-induced yield-stress has also been reported[10,11] in an electrorheological (ER) fluid composed of a colloidal suspension of urea-coated nano-particles of $Ba_{0.8}Rb_{0.4}TiO(C_2O_4)_2$ (U-BRTOCO) in silicone oil, suggesting an unusual field effect on the ε' of nano-structure materials. In the hope that a negative dielectric constant at low frequencies can be found in nano-particle materials, we have therefore investigated the dielectric constant of the ε'(ω) of the U-BRTOCO nano-particle aggregates in the presence of an electric field.

## 4. EXPERIMENTAL AND RESULTS

The complex dielectric constant $\varepsilon(\omega)_{k\neq 0}$ of the U-BRTOCO nano-particle ER fluid samples[11] was measured at room temperature under a dc bias field ($E_b$) between two plate-electrodes. $\varepsilon(\omega) \equiv \varepsilon'(\omega) - i\varepsilon''(\omega) = \varepsilon'(\omega) - i[\sigma(\omega) - \sigma(0)]/\omega = -iI_{ac}d/(\omega V_{ac}S)$, in the Heaviside-Lorentz units, is determined by measuring the ac current $I_{ac}$ through and the ac voltage $V_{ac}$ across the electrodes, with d and S being the distance between the electrodes and the area of the electrode-plates, respectively. The values used are, respectively, $V_{ac}$ = 0.5 – 10 V, $\omega = 10^{-4} – 10^5$ Hz, $E_b$ = 0 - 5.0 kV/mm, d = 0.1 - 0.2 mm, and S = 1cm². The corresponding wave-vector $|\mathbf{k}|$, therefore, is around $10^0$-$10^2$ cm$^{-1}$ if the fluid is treated as a homogeneous medium,[2] but might be as large as $10^5$ cm$^{-1}$ when the nanostructures are considered.

Typical ε'(ω) results of an ER sample at different $E_b$'s are displayed in Fig. 1. At $E_b$ = 0, a large dispersion is observed to fall into two frequency regions (Fig. 1). A Debye-like relaxation appears around $5 \times 10^3$ Hz with the ε' decreasing from ~ 30 to ~ 10 between $10^2$ and $10^5$ Hz. It should be noted that the surface conductance of individual particles has been invoked to account for the similar features previously reported.[12] Below 1 Hz, however, both ε' and ε'' increase as ω decreases, with ε' rapidly reaching a value of 20,000 at ~ $10^{-4}$ Hz. They roughly obey the fractional power law of $\omega^{-n}$ with n ≈ 0.5, and the ε'/ε''-ratio is independent of ω, as prescribed by the empirical "universal dielectric response" (UDR) for a disordered solid, which can be modelled as a capacitor-resistor network.[13] Therefore, in the absence of an $E_b$, our samples behave at low ω similarly to other disordered systems previously reported.

However, in the presence of an $E_b \geq 0.15$ kV/mm, the situation changes drastically: below ~ 1 Hz, the magnitude of ε' decreases rapidly and its sign switches abruptly from positive to negative at a characteristic frequency $\omega_c$ with $\omega_c$ increasing as $E_b$ increases but with a negligible $E_b$-dependent ε' at frequencies above ~ 1 Hz. The electric field effect appears to saturate at $E_b$ > 3.5 kV. As shown in Fig. 1, the positive ε' drops precipitously to a negative value at $\omega_c$ and continues to drop to a more negative value at lower ω. At ω ~ $10^{-3}$ Hz, ε' decreases from $+10^3$ to $-10^4$, when $E_b$ is enhanced from 0 to 0.5 kV/mm, while $\omega_c$ increases from ~$3 \times 10^{-3}$ to $3 \times 10^{-1}$ Hz as $E_b$ rises from 0.5 to 3.5 kV/mm. However, both ε'(ω) and $\omega_c$ change little for $E_b$ > 3.5 kV/mm. As displayed in Fig. 2, a large low frequency dispersion in $\varepsilon''(\omega) = [\sigma(\omega) - \sigma(0)]/\omega$ is also observed. Below a characteristic frequency $\omega_\sigma > \omega_c$, σ(ω) increases with $E_b$ and becomes ω-independent and both $\omega_\sigma$ and $\omega_c$ increase with $E_b$. When $\omega > \omega_\sigma$, σ(ω) is $E_b$-independent (Fig. 2). A slight dip in σ(ω) is detected near $\omega_c$. While the σ(ω) can be understood in terms of the percolation paths associated with a field-induced closer particle-packing, a negative ε'(ω) is not expected from either the percolation or the UDR model.[13]

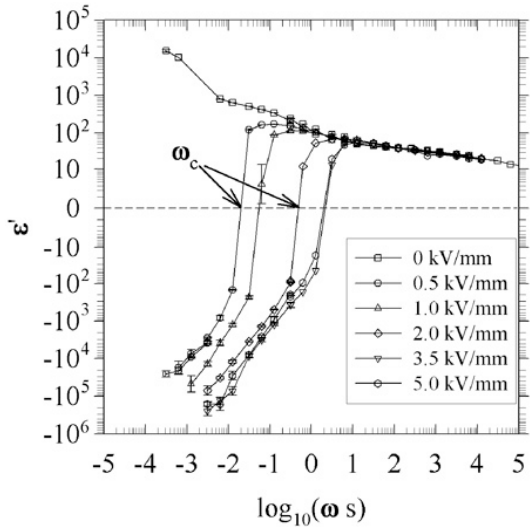

**Figure 1.** The ω – dependence of ε' of U-BRTOCO nano-particles in silicone oil under different $E_b$'s.

To determine the possible effects of the silicone oil, particle-size, particle-material, and temperature on the ε'-sign switching by an electric field, we examine the cold pressed pellet samples of U-BRTOCO nano-particles without silicone oil at different temperatures, and colloids of nano-particles of $Al_2O_3$ of different sizes.[14] Similar field-induced sign switching effect on ε'(ω) is observed in the cold-pressed pellet sample (Fig. 1), demonstrating that the silicone oil is not essential in the manifestation of a negative ε'. Similar field-induced negative ε'(ω) has also been detected in nano-particle colloid of $Al_2O_3$, suggesting that the field-induced sign change may not be so sensitive to materials qualitatively. This led us to investigate the size effect on the sign change. The results are summarized in Fig. 3, which shows that the field-induced ε'-switching clearly depends on the particle size. Down to $10^{-4}$ Hz there is no switching up to 5 kV/mm when the particle size is greater than 0.5 μm; switching is barely detectable up to 15 kV/mm when the size is 11 nm; and ε' drops to $-9.5\times10^2$ at $E_b = 10$ kV/mm when the particle size is 36 nm. The results show that there may exist an optimal particle-size close to $\sim 30$'s nm for ε' of $Al_2O_3$ particles to switch sign by an electric field. It may be interesting to note that at $10^{-4}$ Hz, the ε'-drop to $\sim -10^5$ for U-BRTOCO at $E_b = 3.5$ kV is much greater than for $Al_2O_3$, suggesting that the switching depends on material quantitatively. Changing the temperature from room temperature $\sim 27$ °C to 50 °C enhances ε' above the switching, as well as the $\omega_c$, as shown in Fig. 4. Details are being investigated.

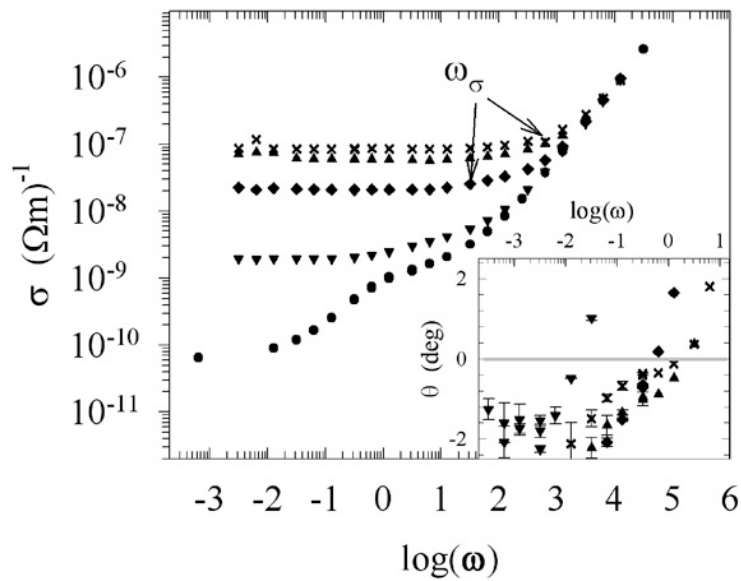

**Figure 2.** The ω – dependence of σ of U-BRTOCO nano-particles in silicone oil under different $E_b$'s. Inset - the phase-angle θ at different ω and $E_b$ with the experimental resolution represented by the width of the horizontal line at θ = 0.

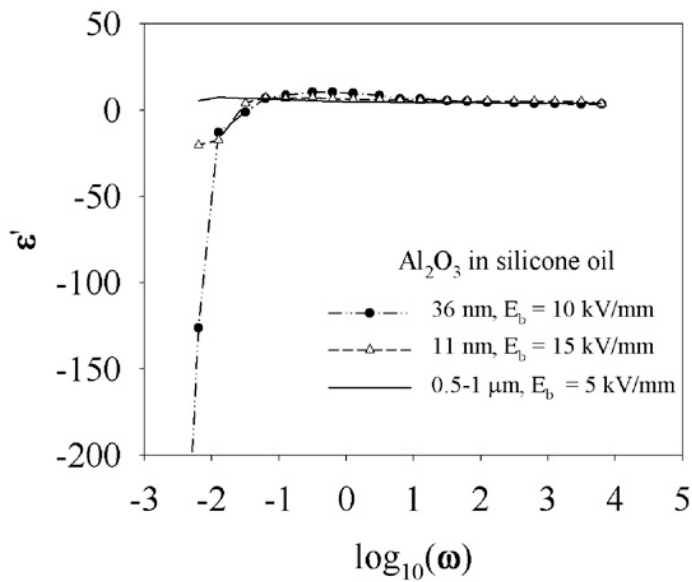

**Figure 3.** Particle size effect on the sign-switching of ε'(ω) of $Al_2O_3$ in silicone oil.

## 5. DISCUSSION

While a field-induced negative ε' at low ω has been reported previously in material systems such as Schottky diodes[15] and fuel cells,[16] it had been considered to be a local electrochemical effect, thermal-induced de-trapping, or simply an experimental artifact. Indeed, to determine ε' at very low ω is a challenge and extra cautions have been taken for the present investigation. A technique-related artifact is not the cause since we employed ten different circuits and all yielded the same results.[17] Calibrations against resistors and capacitors show that both the systematic phase-shifts as well as the experimental resolution of the phase-angle $\theta(\omega)=\tan^{-1}(\omega\varepsilon'/\sigma)$ are better than 0.05°, which is orders of magnitude smaller than our experimental values as shown in the inset of Fig. 2 where the experimental uncertainty is represented by the horizontal line at 0°. The possibility of an electrode-related artifact is unlikely since the same ε'-switching effect is obtained when different electrodes of Cu, Ni, Pt, and Au with different d's are used. Opposite polarities of the applied $E_b$ produce the same negative ε', as displayed in Fig. 5, effectively ruling out any local battery origin of the observation. In addition, the proposed electrochemical reactions and electrode effects should result in a large nonlinearity. Therefore, the possible nonlinear effect was explored[14] by investigating the waveform of the out-of-phase current, the third harmonic, and the $V_{ac}$ and $E_b$-dependences of ε'. The nonlinearity so obtained below $\omega_c$ is not more than a few percent for $V_{ac}$ up to 100 V at $E_b > 2.7$ kV/mm, as demonstrated in Fig. 6. No magnetic field greater than $10^{-6}$ T has been detected in our samples with a negative ε' $\sim -10^5$, ruling out any inductive origin of the phenomenon. It is known that particles of similar charge attract in a material with a negative ε'(ω), which will lead to instabilities such as the formation of charge density waves. However, the observed noise spectrum follows the standard 1/ω–dependence, and displays no anomaly at $\omega_c$, indicating that there exists no electronic instability in the state of negative ε'. The arrest of such instabilities may result from the $E_b$ applied. All of the above demonstrate that the field-induced negative ε' is indeed an intrinsic property of the nano-particle aggregates.

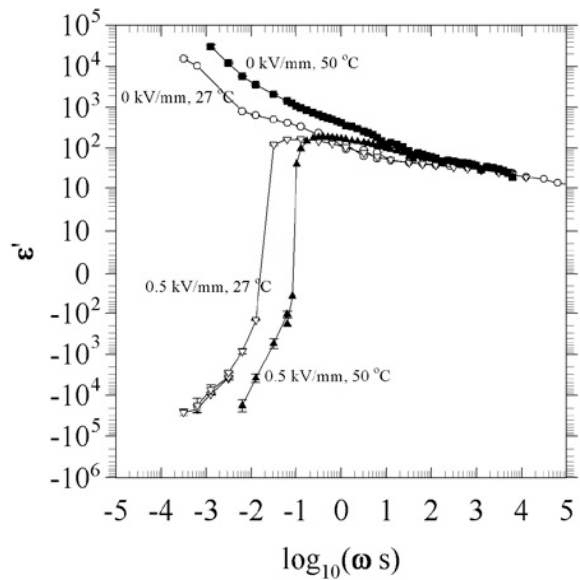

**Figure 4.** The temperature influence on the sign-switching of ε'(ω) of U-BRTOCO nano-particles in silicone oil.

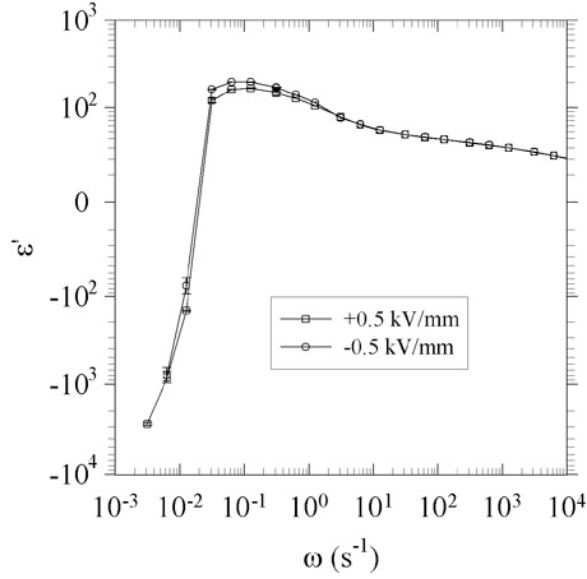

**Figure 5.** Polarity effect on the sign-switching of ε'(ω) of U-BRTOCO nano-particles in silicone oil.

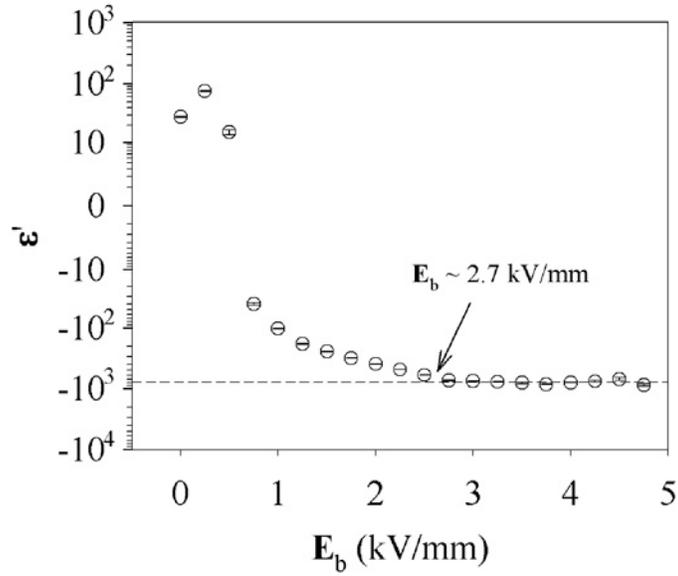

**Figure 6.** The $E_b$ – effect on ε'(ω) of U-BRTOCO nano-particles in silicone oil at $2\times10^{-2}$ Hz.

In a slightly damped system that exhibits a resonance at a frequency $\omega_o$, the equation of state under a driving field of $E_o e^{i\omega t}$ is given by $x'' + \gamma x' + \omega_0^2 x = E_o e^{i\omega t}$, where $\gamma$ is the damping factor. The dielectric constant of the system is then

$$\varepsilon'(\omega) = 1 - \frac{(\omega^2 - \omega_o^2)\omega_p^2}{\omega^2\gamma^2 + (\omega^2 - \omega_o^2)^2}, \qquad (7)$$

where $\omega_p = \sqrt{\dfrac{k}{m}}$ with an elastic modulus $k$ and mass $m$ for a mechanical system or $\sqrt{\dfrac{Ne^2}{m}}$ with a charge carrier density $N$, the particle charge $e$ and the particle mass $m$. A negative $\varepsilon'(\omega)$ occurs at frequencies slightly higher than $\omega_o$. The nano-particles in our ER fluid sample are known[11] to form columns in the presence of an $E_b$. A possible electric-field-driven mechanical resonance at $\omega_o$ in such column-structures is examined. However, the estimated $\omega_o = \sqrt{(k/m)} \sim 10^5$ Hz is much greater than the $\omega_c \sim 10^{-1}$ Hz observed. This is further confirmed by the observation of a negative $\varepsilon'(\omega)$ with a low-$\omega_c$ in the cold-pressed pellet sample, where mechanical oscillation, if it exists, is expected to occur at an even higher frequency. On the other hand, a free plasma[18] gives $\varepsilon'(\omega) = 1 - \dfrac{\omega_p^2}{\gamma^2 + \omega^2}$, which accounts for the data well over more than three orders of magnitude of the negative $\varepsilon'(\omega)$. The $\gamma$ deduced is rather small with a value $\sim 10^{-4}$ s$^{-1}$. The small $\gamma$ implies that the carriers will keep their initial phase intact for hours without significant interference from particle collisions, thermal excitation, and electromagnetic noise. This suggests that new collective excitations associated with interfaces may be the origin of the negative $\varepsilon'(\omega)$. Unfortunately, the $\omega_p \sim 10^{-1}$ Hz in the free plasma model so obtained leads to a very low carrier density of $10^4 - 10^6$/cm$^3$, whose physical meaning remains a puzzle. However, the impasse may be alleviated to a large extent by increasing the effective mass of the carriers. Unfortunately, the observed $\sigma(\omega) = \sigma(0) - \omega\varepsilon''(\omega)$ includes the *dc* percolative conductance and thus prevents a direct comparison with the model.

In an attempt to determine if a negative polarization P exists in the U-BRTOCO nano-particle aggregates in the presence of a positive field, we have measured the I(V)-loops of a sample over a V-range of ±200 V (or field-range of ±2 kV/mm) at frequencies ranging from $10^{-4}$ to $10^0$ Hz. At 1 Hz, a singly connected clockwise loop centering on the origin is detected, as expected of an ordinary capacitor. However, at lower $\omega$, two counter-clockwise loops appear at high |V| (inset, Fig. 7), in addition to the clockwise one around the origin (too small to be resolved with the scale used in the inset of Fig. 7), suggesting the reversal of the I-sign at high |V|. To investigate the charge Q accumulated across the sample, we have integrated the hysteretic component of I, i.e. $Q(t) = \int_0^t (I - I_{reversible}) d\tau$, which is proportional to P and shown together with the V as a function of time in Fig. 7. Q or P is found to be negative under a positive V, except for a narrow V-range within ± 15 V. The negative $\varepsilon'$ and the induced Q or P opposing the applied field in the U-BRCOTO nano-particle suspension in silicone oil appear to be the expected properties of the missing "diaelectricity," the dielectric analogue of a diamagnet. Unusual electrophoretic effects are also observed under an optical microscope, and shown in Fig. 8, when an electric field is applied between the two transparent electrodes. The U-BRCOTO nanoparticles appear to be pushed out from the electric field where two electrodes overlap, although some nano-particles are left behind to form some kind of pattern, reminiscent of the magnetic field expulsion from a Type II superconductor. However, it should be pointed out that the length scales in the two cases are drastically different.

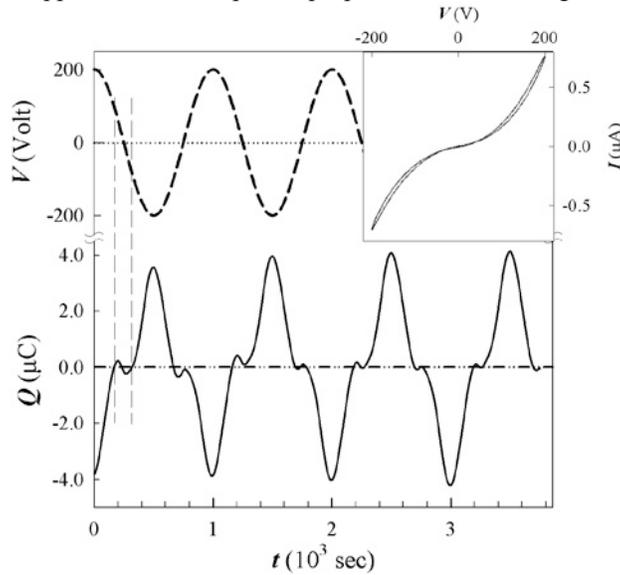

**Figure 7.** The t – dependences of V and Q of U-BRTOCO nano-particles in silicone oil. Inset – the I-V loop at $\omega < 1$ Hz.

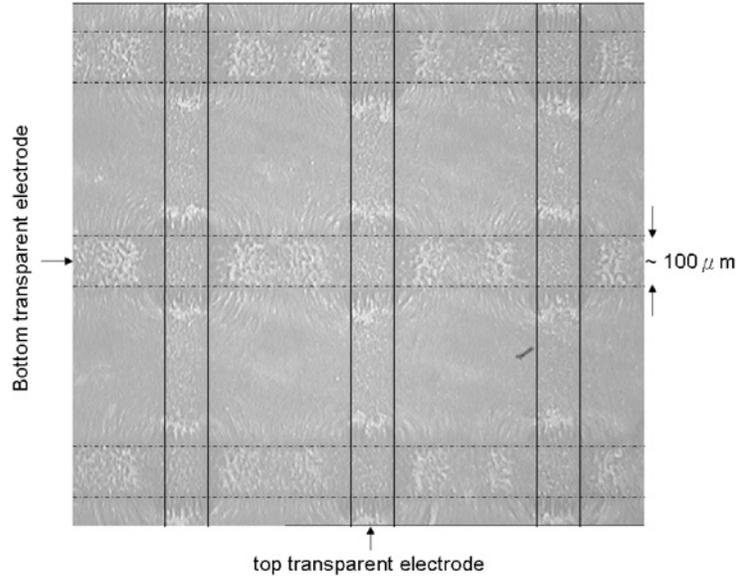

**Figure 8.** The reflective optical image of U-BRTOCO nano-particles in silicone oil under an $E_b$ = 2 kV/mm.

## 6. CONCLUSION

We have measured at room temperature (~ 27 °C) the dielectric constant $\varepsilon'(\omega)$ of samples of U-BRTCOTO and of $Al_2O_3$ nano-particles of different sizes suspended in silicone oil or cold-pressed U-BRTCOCO nano-particles, whose dielectric behavior in the absence of an electric field is typical of samples of nano-particles previously studied. However, $\varepsilon'(\omega)$ is drastically changed by the application of an electric biased field $E_b \geq 0.15$ kV/mm below ~ 1 Hz but not above, *e.g.* $\varepsilon'$ switches sign from positive to negative at a characteristic frequency $\omega_c$ which increases with $E_b$ and undergoes a large change in its value (for instance from $+10^3$ to $-10^4$ at $10^{-3}$ Hz with $E_b$ = 0.5 kV/mm). On the other hand, the $\varepsilon'(\omega)$ – behavior stops changing when $E_b$ exceeds 3.5 kV/mm. Tests carried out show that the observations represent the intrinsic properties of the sample and cannot be associated with the electrode or with a local chemical or inductive effect. The field-induced sign-switching of $\varepsilon'(\omega)$ below ~ 1 Hz depends on the size of the particles with a possible optimal value ~ 30's nm but less so on the material of the particles. A charge (or polarization) opposite to the applied voltage and an unusual electrophoretic effect of the nano-particles in silicone oil are also detected, reminiscent of the characteristics of the missing "diaelectricity." The negative $\varepsilon'(\omega)$ observed can be accounted well over three orders of magnitude by the free plasma model with two fitting parameters, namely carrier concentration and the damping factor. The small fitting parameters so obtained suggest that the yet-to-be-determined collective excitations of large effective mass in the interfaces induced by electric field may be responsible for the negative $\varepsilon'(\omega)$ of the nano-particle aggregates. These proposed collective modes may lead to novel physical states with extremely small damping such as possible high temperature superconductivity. The unusual electrophoretic effect detected needs further elucidation. Creating a controlled system to simulate the nano-particle aggregates for a definitive characterization of the observations is under way.

## ACKNOWLEDGEMENT

The work in Houston is supported in part by AFOSR Award No. A9550-05-1-0447, NSF Grant No. DMR-9804325, the T. L. L. Temple Foundation, the John J. and Rebecca Moores Endowment, the Strategic Partnership for Research in